\documentclass[nato,noid,numreferences]{crckbked}

\usepackage{graphicx}
\usepackage{epsf}

\def\startappendices{%
\appendix
\renewcommand{\thesection}{\Roman{section}}
\setcounter{section}{0}
\setcounter{subsection}{0}
\setcounter{subsubsection}{0}
}

\def\d{{\rm d}}

\def\tr{\rm tr\:}

\begin{document}

\begin{opening}
\title{Monopoles from Instantons
\vskip-2cm\hfill{\rm INLO-PUB-02/02}\vskip1.4cm}

\author{Falk Bruckmann}
\institute{Instituut-Lorentz for Theoretical Physics, Leiden University\\
           P.O. Box 9506, NL-2300 RA Leiden, The Netherlands\\
           {\tt bruckmann@lorentz.leidenuniv.nl}}

\begin{abstract}
The relation between defects of Abelian gauges and instantons is
discussed for explicit examples in the Laplacian Abelian gauge. 
The defect coming from an instanton is pointlike and becomes
a monopole loop with twist upon perturbation.
The interplay between magnetic charge, twist and instanton number --
encoded as a Hopf invariant -- is investigated
with the help of a new method, an auxiliary Abelian fibre bundle.
\end{abstract}

\renewcommand{\thefootnote}{\fnsymbol{footnote}}
\footnotetext[1]{
Talk given at the NATO Advanced Research Workshop on {\sl Confinement,
Topology, and Other Non-Perturbative Aspects of QCD},
January 21-27, 2002, Star\'a Lesn\'a, Slovakia.}
\renewcommand{\thefootnote}{\arabic{footnote}}

\end{opening}

\section{Introduction}

In order to explain confinement in quantum chromodynamics, the
{\it dual superconductor} scenario was proposed long ago. It
states that the condensation of magnetic monopoles forces the
chromoelectric flux into tubes. This results in a linear potential
confining the quarks.
Analogously, large Wilson loops in pure Yang-Mills theories fall off
with an area law.

However, monopoles can be identified in such a theory only after
special gauge fixings, named Abelian gauges \cite{thooft:81a}.
 They are best described
by the diagonalisation of an auxiliary Higgs field $\phi$ in the
adjoint representation. For gauge group $SU(2)$ -- to which we will
restrict ourselves in the following -- this is tantamount to bringing
$\phi$ into the third color direction. The residual gauge freedom then
consists of rotations around this axis. It constitutes the Abelian gauge
group $U(1)$ (of diagonal matrices in $SU(2)$). The Abelian gauge is
ambiguous at points in space-time where the field $\phi$ vanishes.
These so-called defects are generically (closed)
lines in four
dimensions. Around them one can smoothly define the normalised Higgs
field $n=\phi/|\phi|$. The latter is a mapping from an $S^2$ in
coordinate space to another $S^2$ in color space representing the
coset $SU(2)/U(1)$. Generically, the field $n$ is a hedgehog
around each defect with winding number $\pm 1$.
Therefore, its diagonalisation 
leads to a Dirac monopole with appropriate magnetic charge.

The dual superconductor picture is supported by a number of lattice
tests. On the other hand, it has to face the fundamental problem of
gauge dependence. In this context, the {\it relation of defects to
instantons} is of interest, because the notion of instantons
(or, more generally, configurations with instanton number) is
a gauge-independent one. A physical motivation is the fact that
instantons 
are responsible for other effects like chiral symmetry
breaking. More technically, both the magnetic charge and the instanton
number are topological quantities. 

In the course of the talk I will present explicit examples of
defects/mono\-poles coming from instantons: the single instanton in the
Laplacian Abelian gauge induces a pointlike defect, which becomes a
monopole loop after a deformation. At the same time I will discuss the
topological properties of these objects: the instanton number is
translated into the Hopf invariant of $n$ via transition
functions/boundary conditions. The monopole loop generates this Hopf
invariant by virtue of magnetic charge plus twist. 
This known statement will be shown by a new 
method, an auxiliary
Abelian fibre bundle. Being of topological origin, these
considerations are then valid for any configuration in any Abelian gauge. 

\section{Point defect from instanton}

The Higgs field of the Laplacian Abelian gauge is defined as the
ground state of the gauge covariant Laplacian \cite{vandersijs:97}.
 It is a popular
Abelian gauge on the lattice. In order to obtain a discrete spectrum
for the Laplacian in the continuum, one better works on a finite volume
manifold like the four-sphere\footnote{Conformal
invariance can be used to write
down the instanton configurations on the geometrical $S^4$.}.

The ground state of the Laplacian in the background of a single
instanton has been solved in \cite{bruckmann:01a}. The high symmetry
of the background allows for an analytic solution, which however turns
out to be non-generic:
the Higgs field vanishes quadratically at the instanton position. Such
a {\it point defect} has been observed for instantons on the lattice, too
\cite{deforcrand:xx}. The field $n$ is now a mapping from an $S^3$
around the defect to $S^2$. These mappings are characterised by
another integer, the Hopf invariant, $H(n)\in\pi_3(S^2)\cong\mathbb{Z}$.
It is briefly discussed in the appendix.

\begin{figure}[b]
\centering
\begin{minipage}{0.49\linewidth}
\epsfxsize=6.2cm\epsffile{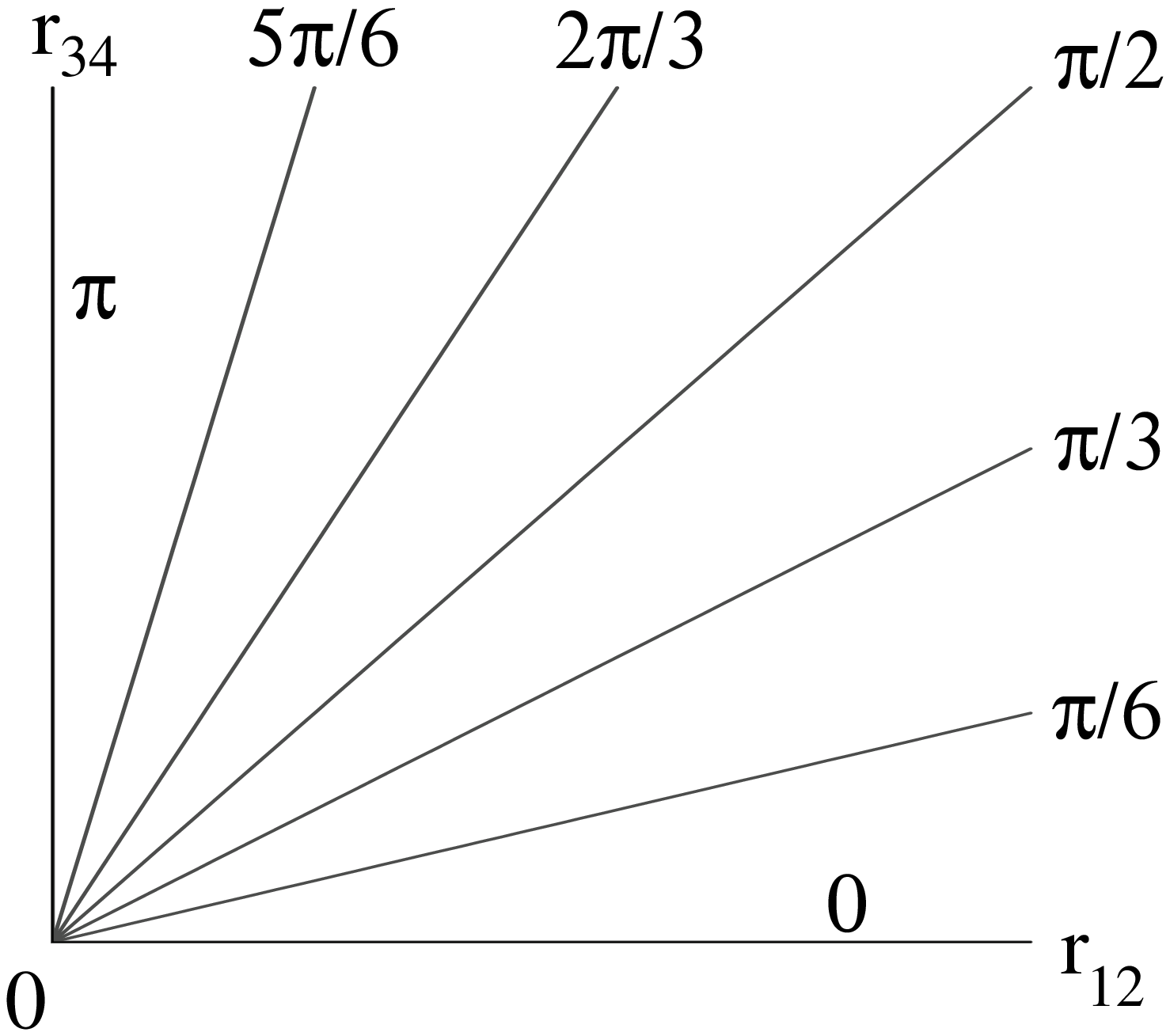}
\end{minipage}\hfill
\begin{minipage}{0.49\linewidth}
\epsfxsize=6.2cm\epsffile{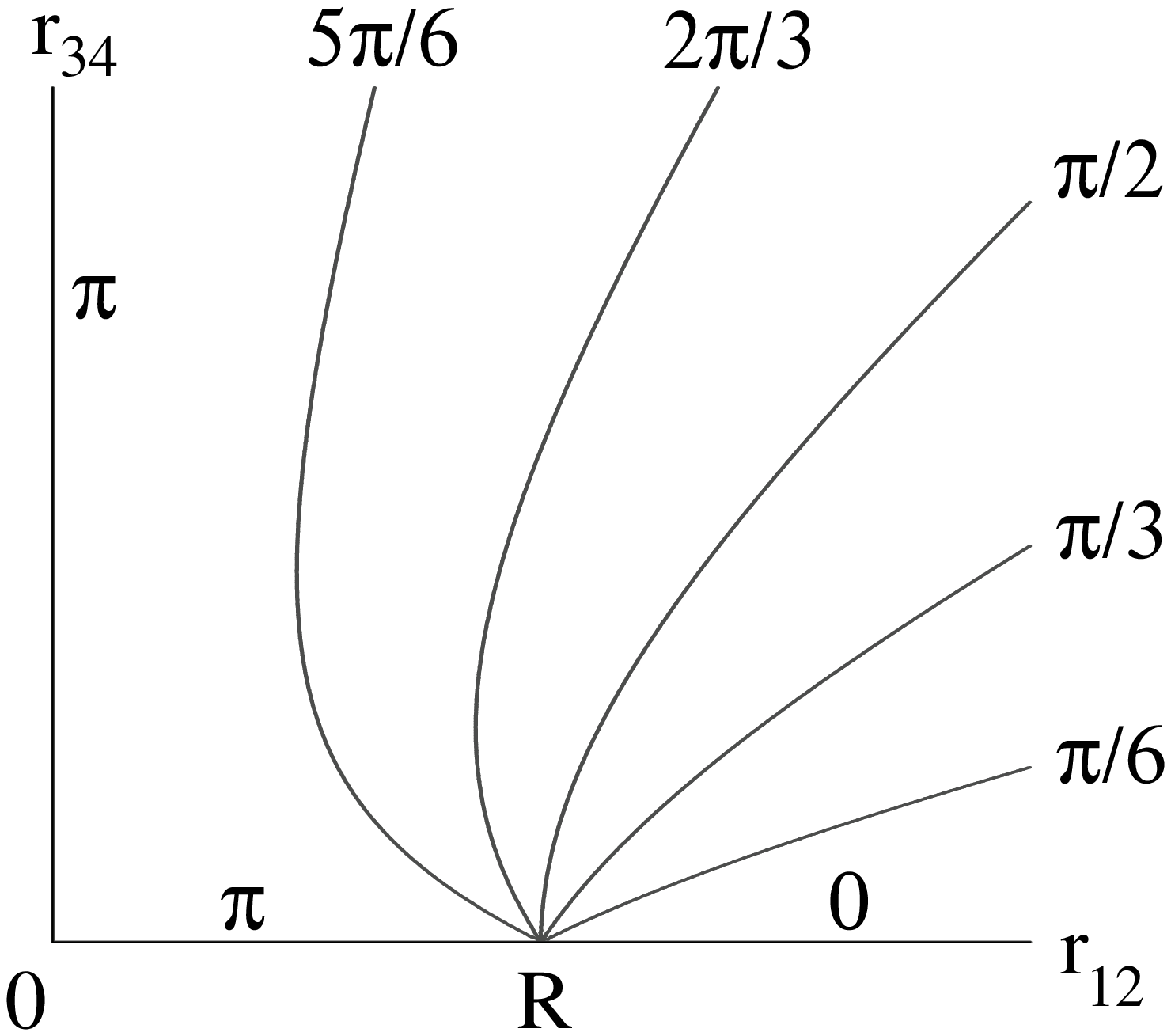}
\end{minipage}
\caption{The normalised Higgs field $n$ of the point defect (left) and
the monopole loop (right), i.e. before and after deformation. Shown
are lines of constant azimuthal angle of $n$ with 0 and $\pi$ standing
for the the north and south pole, respectively. Defects are
singularities in $n$, where those lines meet. The
four-dimensional space-time is visualised by radii
$r_{12}^2=x_1^2+x_2^2,\: r_{34}^2=x_3^2+x_4^2$.}
\label{isolines}
\end{figure}

For the example at hand the $n$-field is
the `standard Hopf map' sketched in Figure \ref{isolines}. 
It has Hopf invariant 1, which is not a coincidence as becomes clear in
the next section.

\section{Hopf invariant and instanton number}

The topology of the Higgs field is governed by the demand of gauge
fixing that gauge
equivalent configurations shall be gauged to the same
configuration\footnote{up to a residual Abelian gauge transformation,
which does not transform the coset field $n$}. Therefore, the Higgs
field must have the same transition functions/boundary conditions as
the gauge field. This means in particular, that $n$ has a Hopf
invariant equal to the instanton number \cite{jahn:00}. Both of these
quantities are measured on a large three-sphere, the transition
region/the boundary of space-time.

For the point defect the Hopf
invariant appears already at the little sphere since there is no other
defect present.

\section{Monopole loop as deformation}
\label{monopol_section}

For configurations near the instanton in configuration
space, 
$A=A_{\rm inst}+\delta A$, one can in principle perform a perturbation
theory $\phi=\phi_{\rm inst}+\delta\phi$ \`a la Schr\"odinger. Due to
lack of knowledge of the full spectrum of the Laplacian this has not
been solved. Nevertheless, to investigate the behaviour of the defect,
it is sufficient to Taylor expand the perturbation $\delta\phi$ at
the defect \cite{bruckmann:01b}. For our purposes even the zeroth
order will be enough. Without loss of generality we specialise to a
perturbation in the third color direction. So finally
$\delta\phi=R^2\tau_3$ since the Higgs field is of dimension
(length$)^2$.

A straightforward calculation shows that upon this perturbation the
defect becomes a monopole loop, namely a circle of radius $R$ in the
$x_1x_2$-plane. 
This configuration is now generic, the perturbation has (partly) broken the
symmetry. 
The field $n$ with its typical hedgehog behaviour is
depicted in Figure \ref{isolines}. 
It also shows the local nature of the perturbation. 
The Higgs field is not changed at large distances
and still has Hopf invariant 1.
Actually the configuration is the one proposed in \cite{brower:97b}
for the Maximally Abelian gauge, but there it is suppressed by
the gauge fixing functional. 

What is the precise relation between the monopole loop in the bulk and
the Hopf invariant = instanton number at the boundary?
It has to do with the twist\footnote{also called Taubes winding}
\cite{taubes:84a} meaning the {\it rotation of the Higgs field around an
axis in color space while moving along the loop}. For our case, the
polar angle of $n$ depends on the worldline coordinate
$\varphi_{12}=\arctan(x_2/x_1)$ and performs just one rotation. Beside
the twist, there is the possibility that the instanton number gets
contributions from a two-dimensional sheet spanned by the monopole
loop \cite{jahn:00}. This can be thought of as the set of Dirac
strings. The latter will not show up when we discuss the interplay
between magnetic charge, twist and instanton number with the help of
an auxiliary Abelian fibre bundle in the next section.

\section{Auxiliary Abelian fibre bundle}

The idea of this section is to describe the properties of the Higgs
field $n$ in terms of an auxiliary Abelian gauge field $a$ and its
field strength $f=\d a$. These will have immediate physical
interpretations.
Using the diagonalising gauge
transformation of $n$ --  the one which transforms the non-Abelian gauge field
into the Abelian gauge -- these fields have the following expressions,
\begin{eqnarray*}
a&=&(ig\d g^\dagger,\tau_3)\,,\qquad\quad g n g^\dagger=\tau_3\,,\\
f&=&(g\d g^\dagger,\tau_1)\wedge(g\d g^\dagger,\tau_2)=(in,\d n\wedge\d n)\,.
\end{eqnarray*}
$a$ is the Abelian projected inhomogeneous part of the gauge
transformed non-Abelian field. $f$ is the topological density of $n$,
its integral over a two-sphere gives the winding number = magnetic
charge.
Therefore, these objects carry information about the monopoles in the
bulk.
On the other hand, they are the ingredients of the Abelian Chern-Simons form
$CS(a)=a\wedge f/16\pi^2$ which gives the Hopf invariant.

The only subtlety is that $g$ and thus $a$ may not be defined globally
(due to topological obstructions) even if $n$ and thus $f$
are. Accordingly, one has to work with patches in the framework of a bundle.

\begin{figure}[b]
\centering
\begin{minipage}{0.45\linewidth}
\epsfxsize=5.0cm\epsffile{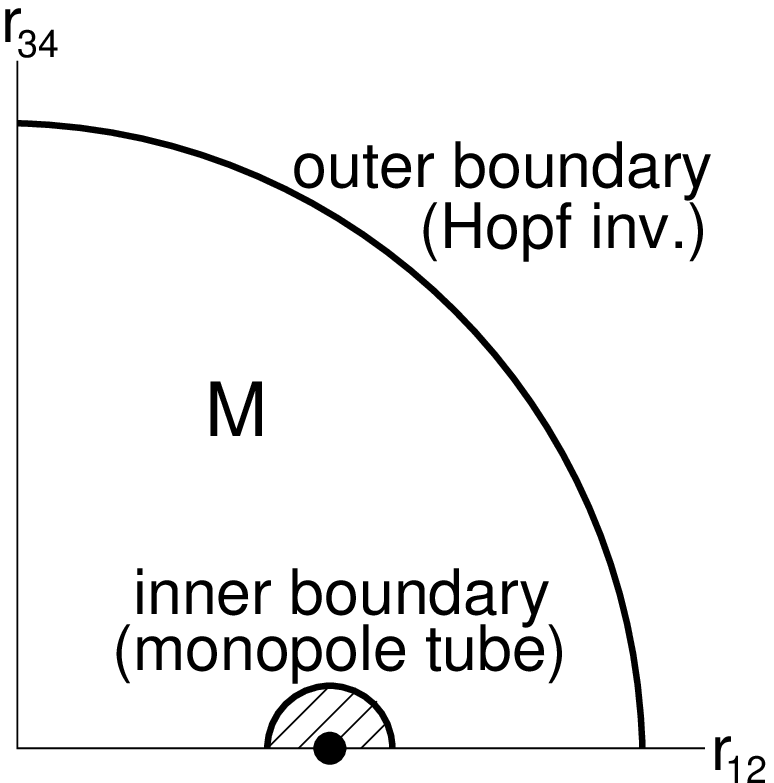}
\end{minipage}\hfill
\begin{minipage}{0.45\linewidth}
\epsfxsize=5.0cm\epsffile{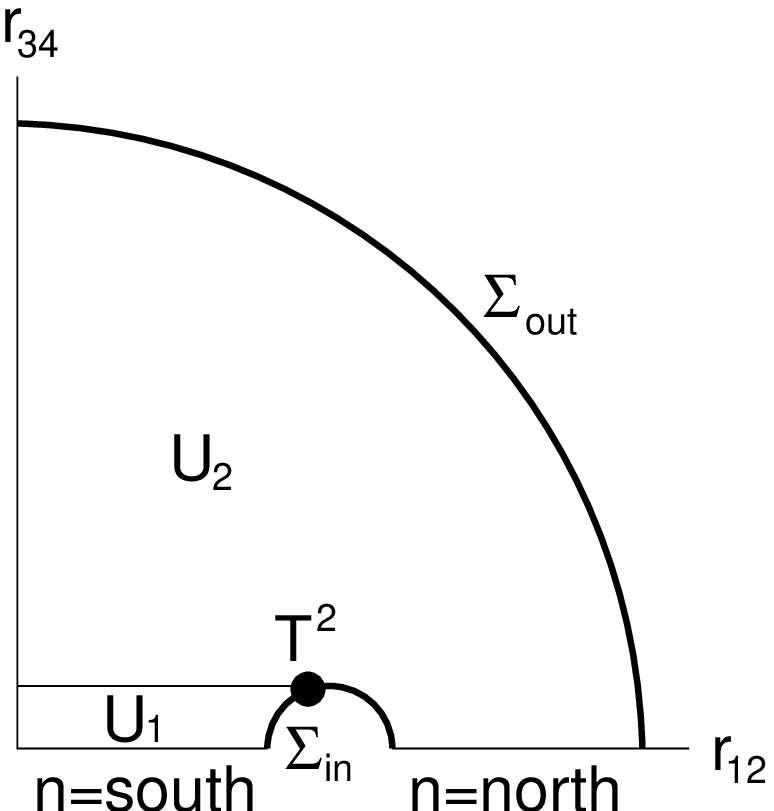}
\end{minipage}
\caption{The construction of the base manifold $M$ by cutting out the
monopole loop (and infinity) and its minimal patching with
$U_1$ and $U_2$. The contribution of the monopole loop to the
(vanishing) instanton number reduces finally to the shown two-torus.}
\label{base}
\end{figure}

For the base manifold $M$ of this bundle, i.e.~the space-time, we have to
cut out a tube surrounding the monopole loop, since $n$ is not
continuous there. When working on $S^4$ one also has to cut out the
pole which represents infinity on $\mathbb{R}^4$, the  transition
functions there are genuinely non-Abelian.
So, Abelianisation has lead to a space-time $M$ 
with two boundaries (cf. Figure \ref{base}).
The outer boundary $\Sigma_{\rm
out}=S^3_\infty$ carries information about the Hopf invariant, the
inner boundary $\Sigma_{\rm in}=S^2\times S^1$ contains the monopole
loop.

$n$ is smooth over $M$ by definition. Its diagonalisation will be
smooth only over patches $U_i$, $g_i n g^\dagger_i=\tau_3$.
The local gauge field reads $a_i=(ig_i\d g^\dagger_i,\tau_3)$. Indeed,
the residual gauge freedom $g\rightarrow\exp(i\lambda\tau_3)g$
transforms it as $a\rightarrow a+\d\lambda$. In particular the
transition on the overlap of two patches is $a_i=a_j+\d\lambda_{ij}$.

The key point in the construction is the vanishing of the topological
density $f\wedge f$ due to its form content\footnote{$n$ has two
degrees of freedom and can at most constitute a two-form, namely
$f$.}. 
As will be explained now, the zero instanton number of the auxiliary field
interpolates between the boundaries. In a way, our method is similar
to the residue calculus.

It is well-known, that the instanton number can be expressed in terms
of transition functions $\Omega_{ij}$ \cite{vanbaal:82}.
For the case of a
non-Abelian theory without boundaries the formula reads,
\begin{eqnarray*}
\nu(A)=\frac{1}{48\pi^2}\sum_{i,j}\int_{U_i\cap U_j}\!\!\!\!
            \tr(\Omega_{ij}^\dagger\d \Omega_{ij})^3
      +\frac{1}{48\pi^2}\sum_{i,j,k}\int_{U_i\cap U_j\cap U_k}\!\!\!\!
            \tr(\Omega_{ij}\d \Omega_{ij}^\dagger\wedge
               \d\Omega_{jk}^\dagger\Omega_{jk})\,,
\end{eqnarray*}
where we have suppressed orientations.
Abelian transition functions can only contribute to the second
term. We note in passing that for Abelian instantons on the
four-torus  these are integrals over two-tori.

For a manifold with boundaries one has to include two more terms,
\begin{eqnarray*}
0=\nu(a)=&&\!\!\!\!\!
-\frac{1}{96\pi^2}\sum_{i,j,k}\int_{U_i\cap U_j\cap U_k}
\d\lambda_{ij}\wedge\d\lambda_{jk}\\
&&\!\!\!\!\!+\frac{1}{16\pi^2}\sum_{i}\int_{U_i\cap \Sigma}a_i\wedge f
+\frac{1}{32\pi^2}\sum_{i,j}
\int_{U_i\cap U_j\cap \Sigma}a_i\wedge\d\lambda_{ij}\,.
\end{eqnarray*}
The new terms are
boundary contributions with a similar structure, but not entirely
given in terms of transition functions. They will be interpreted in
terms of the Hopf invariant and monopole properties. The old term is not
of this form but only occurs for more than two patches. Therefore we
continue with the search for a minimal patching of $M$.

How big are the subspaces $U_i$ of $M$ on which on can define a smooth
diagonalisation $g_i$? In other words, given a smooth field strength
$f$ fulfilling the Bianchi identity $d f=0$, on which $U_i$ can one
define smooth $a_i$ with $f=\d a_i$? The answer is the second
cohomology $H^2(U_i)$. Loosely speaking, the patches $U_i$ need not be
contractable but are not allowed to have two-dimensional holes.

For a strip near the inner boundary we have $H^2(S^2\times
S^1\times\mbox{interval})\neq 0$. The $S^2$-factor gives the magnetic
charge. This $S^2$ needs to be divided into two patches for every
point on the worldline $S^1$.
This fact is well-known from the Wu-Yang
construction of the Dirac monopole avoiding Dirac strings.
One of the patches should contain the
points where $n$ points say southwards. 

For a strip near the outer boundary we have
$H^2(S^3\times\mbox{interval})=0$. Therefore this volume can be
covered with just one patch, a fact that plays a role in the
definition of the Hopf invariant.

One can match the patches discussed so far in such a manner,
 that the minimal atlas of
$M$ consists of two patches, see Figure \ref{base}.
Firstly, extend the south patch of 
the monopole to a thick sheet $U_1$ spanned by the loop. 
Secondly, glue the other patch $U_2$ of the monopole with the patch at the
outer boundary. In this way the instanton number will not pick up the
mentioned bulk contributions and the instanton topology is
`localised' on the boundary of the monopole tube.

The final step will be the interpretation of the boundary
contributions. This is very easy at the outer boundary. Since it is
covered by one patch, only the second term in the above formula will
contribute. Moreover, it perfectly coincides with the Hopf invariant.
At the inner boundary the second term can be further reduced due to
the third cohomology\footnote{$a_i\wedge f$ is closed and
$H^3(\mbox{interval}^2\times S^1)=0$.}. The whole contribution of this
boundary reduces to a two-torus $U_1\cap U_2\cap\Sigma_{\rm in}$ like
for the Abelian instantons. Moreover, the integral factorises into two
integrals over circles. The first one is the integral of
$\d\lambda_{12}$ over an $S^1$ {\it around} the loop which is nothing but
the magnetic charge. The second one is an integral of the gradient of
the polar angle of $n$ over an $S^1$ {\it along} the loop. In Section
\ref{monopol_section} we have identified this as the twist.
Thus we arrive at the simple formula
\begin{eqnarray*}
\framebox{instanton number = Hopf invariant = magnetic charge $\times$ twist}
\end{eqnarray*}
which is a special case of the complicated one discussed in
\cite{jahn:00}.
For the case at hand we have $1=1=1\times 1$.

\section{Summary and Outlook}

We have discussed the defects induced by instantons in the Laplacian
Abelian gauge. The single instanton leads to a non-generic point
defect. The latter is a seed for a monopole loop (or a monopole loop
`shrunken to zero radius'). A perturbation of the background makes
this monopole visible.
In addition, the monopole loop comes with a
twist. It can be viewed as generating the electric field needed for
the instanton number.
We emphasize that details of the monopole loop like its position
depend on the chosen Abelian gauge.

The topological considerations, however, are gauge-independent.
The instanton number is converted into the Hopf invariant of the
normalised Higgs field $n$. Measured on the boundary, it can be
related to the properties of the monopole in the bulk. For this
purpose we have introduced an auxiliary Abelian fibre bundle. It
avoids singularities like Dirac strings. The task in this approach is
to find a (minimal) patching and to compute and
interpret the remaining boundary
terms. In a very transparent way this leads to the instanton number
being the product of magnetic charge and twist. This method is valid
for any configuration in any Abelian gauge. In principle it can also
be applied to other space-times like tori,
where however the patching will be different.  
The twist has also been found for the constituent monopoles of
calorons.

Topology can only describe the kinematics of monopoles. One might
still speculate, that the shown correlation between instantons and
monopoles\footnote{which for the given examples is even local: the
defects are centered at the instanton core.}
may have implications for the dynamics. This will certainly
be the case when modelling the QCD vacuum as an instanton ensemble.
The instanton liquid has been successful for chiral symmetry
breaking. However, confinement could not be explained by a
reasonable instanton ensemble so far.
It has been observed that the individual monopole loops of two and
more instantons are able to fuse to one larger one. This suggests that
under certain circumstances a percolating monopole loop can be
generated in this way.

\section{Acknowledgements}

The author thanks the organisers for a stimulating workshop in a nice
surrounding. Furthermore he is grateful to
Alexander Bais,
Philippe de Forcrand,
Michael Engelhardt,
Chris Ford,
Oliver Jahn and
Pierre van Baal
for helpful discussions. This work was supported by FOM.

\startappendices

We use half the Pauli matrices $\tau_a=\sigma_a/2$ as the basis in
color space, i.e. as the generators of the Lie algebra $su(2)$. The
scalar product of vectors is $(x,y)=2\,\tr xy=x_ay_a$,
where $x=x_a\tau_a$. 
The wedge product of `colored differential forms' is defined
accordingly $x\wedge y=i/2\,\epsilon_{abc}\,x_a\wedge y_b\,\tau_c$.

The Hopf invariant can be defined as a linking number.
An algebraic definition starts with a two-form $f=(in,\d n\wedge\d
n)$.
Because $f$ is closed and the second cohomology of $S^3$ vanishes,
there exists a one-form
$a$ with $\d a=f$. Together, these forms built a three-form to
integrate over, and $H(n)=\int_{S^3}a\wedge f/16\pi^2$.

This construction can be rewritten in terms of the diagonalising gauge
transformation, cf. Section 5.
The Hopf invariant $H(n)=\int_{S^3}\tr(g\d g^\dagger)^3/24\pi^2$
becomes the winding number of $g$, in the usual sense of a mapping
from $S^3$ to $SU(2)\cong S^3$.


\begin{thebibliography}{00}

\bibitem{brower:97b}
Brower, R.~C., Orginos, K.~N. and Tan, C.-I. (1997)
Magnetic monopole loop for
  the Yang-Mills instanton,
{\it Phys.~Rev.}, {\bf D55}, pp. 6313--6326

\bibitem{bruckmann:01b}
Bruckmann, F. (2001) Hopf defects as seeds for monopole loops,
{\it J.~High Energy Phys.}, {\bf 0108}, p. 30

\bibitem{bruckmann:01a}
Bruckmann, F., Heinzl, T., Vekua, T. and Wipf, A.  (2001) Magnetic Monopoles
  vs.~Hopf Defects in the Laplacian (Abelian) Gauge,
{\it  Nucl.~Phys.}, {\bf B593}, pp. 545--561

\bibitem{deforcrand:xx}
de~Forcrand, P., private communication.

\bibitem{jahn:00}
Jahn, O.  (2000) Instantons and monopoles in general Abelian gauges,
{\it J.~Phys.}, {\bf A33}, pp. 2997--3019

\bibitem{thooft:81a}
't~Hooft, G. (1981) Topology of the Gauge Condition and New Confinement Phases
  in Non-Abelian Gauge Theories,
{\it Nucl.~Phys.}, {\bf B190}, pp. 455

\bibitem{taubes:84a}
Taubes, C.~H. (1984) Morse theory and monopoles: Topology in long-ranged forces,
in G.~'t~Hooft (ed.) {\it Progress in gauge field theory},
Plenum Press, New York 

\bibitem{vanbaal:82}
van Baal, P. (1982) Some Results for SU(N) Gauge Fields on the Hypertorus,
{\it Commun.~Math.~Phys.}, {\bf 85}, pp. 529

\bibitem{vandersijs:97}
van~der Sijs, A.~J. (1997) Laplacian Abelian projection,
{\it Nucl.~Phys.~B (Proc.~Suppl.)}, {\bf 53}, pp. 535--537

\end{thebibliography}
\end{document}